\documentclass[aps,prd,preprint,floatfix,nofootinbib]{revtex4}
\usepackage{graphicx}

\newcommand{\roughly}[1]
    {{\mathrel{\raise.3ex\hbox{$#1$\kern-.75em\lower1ex\hbox{$\sim$}}}}}

\begin{document}

\title{Phenomenology of iQuarkonium}

\renewcommand{\thefootnote}{\fnsymbol{footnote}}

\author{ 
Kingman Cheung$^{1,2}$, 
Wai-Yee Keung$^3$
and Tzu-Chiang Yuan$^{4,5}$
 }
\affiliation{$^1$Department of Physics, National Tsing Hua University, 
Hsinchu 300, Taiwan
\\
$^2$Physics Division, National Center for Theoretical Sciences,
Hsinchu 300, Taiwan
\\
$^3$Department of Physics, University of Illinois, Chicago IL 60607-7059, USA
\\
$^4$Institute of Physics, Academia Sinica, Nankang, Taipei 11529, Taiwan
\\
$^5$Kavli Institute for Theoretical Physics China, CAS, Beijing 100190, China
}

\renewcommand{\thefootnote}{\arabic{footnote}}
\date{\today}

\begin{abstract}
Phenomenology of uncolored iquarks -- hypothetical fermions charged
under a new confining unbroken non-abelian gauge group as well as the
standard electroweak gauge group -- is investigated for the iquark
mass in the range near and above 100 GeV.
If the new confining scale turns out to be higher than MeV but much
less than the iquark mass, the iquark-antiiquark pair produced in the
collider will promptly relaxed into the ground state of the
iquarkonium. Subsequent pair annihilation into standard model
particles gives useful information of the iquark dynamics.
We formulate in details production and decays of the
iquark-antiiquark bound states.
Decay patterns of the iquarkonium can be distinguished
from the superheavy quarkonium 
of a sequential fourth generation of quarks with degenerate mass.

\end{abstract}


\maketitle
 
\section{Introduction}

A new confining unbroken non-abelian gauge interaction, in mimic of
quantum chromodynamics (QCD) of strong interaction, may exist in
models beyond the standard model (SM).  Okun \cite{thetons} pioneered
in this theoretical curiosity long time ago and named these new
particles ``thetons'', including $\theta$-strings, $\theta$-leptons,
$\theta$-quarks, $\theta$-hadrons, etc associated with the gauge group
$SU_\theta(N)$.
This theory can also be regarded as a certain limit of QCD with light
quarks removed such that the scale $\Lambda$ where QCD gets strong is
much smaller than the heavy quark masses \cite{bj,Gupta-Quinn}.
Phenomenology of this hypothetical QCD can be drastically different
from the real world where light quarks do exist and one must have to
worry about the spontaneous chiral symmetry breaking.
For instance, heavy quarks are connected by unbreakable long and stable
string flux tube since there is no light quark pairs popped up from
the vacuum as the heavy quarks are being pulled apart.
However, the $\theta$-fermions $Q$ and $\overline Q$ of the new gauge
interaction, if also carry standard model quantum numbers, have to be
massive ($M_Q \roughly>$ 100 GeV). Otherwise, they would have been
observed in current collider experiments.

These heavy $\theta$-fermions were recently revived and renamed as 
``quirks'' by
Kang and Luty \cite{Kang-Luty}, who emphasized their fantastic event
structures due to the long and stable string flux tube connecting them.
We will briefly mention some of their observations here.
For definiteness, let us assume that this new gauge group is 
$SU_{C^\prime}(N_{\rm IC})$ with its characteristic scale $\Lambda' \ll M_Q$.
The size of the string flux tube is of order ${\Lambda'}^{-1}$
 and it  
can be macroscopic for 100 eV $\roughly< \  \Lambda'  \roughly< $ 10 keV,
or mesoscopic  for 10 keV $\roughly< \  \Lambda'  \roughly< $ MeV, or
microscopic for MeV $\roughly< \  \Lambda'$. 
In general, one also assume $\Lambda'$ is smaller than the 
$\Lambda_{\rm QCD}$
such that the new color degree of freedom bears the name infracolor (IC).
We call this infracolor (a.k.a. iQCD) gluon fields igluons, and
the fermions iquarks.
\footnote{
Kang and Luty \cite{Kang-Luty} called this quirk. To some extent it is 
easy to mix up ``quark'' and ``quirk'' when listening to 
talks of quirks.
So we introduce ``iquark'' (pronounced as i-quark), and similarly
``igluon'' (pronounced as i-gluon), etc.
}
It has been noticed that the infracolor is an example of the ``hidden
valley'' model in Ref.\cite{Strassler:2006im}.
The igluons do not carry SM quantum numbers. Thus,
iglueballs can only couple to SM gauge bosons through a heavy iquark loop
assuming the heavy iquark is not a SM singlet. 
At two-loop level, the iglueballs can also couple to the SM fermions.
Thus, these iglueballs may have escaped detection.
In the macroscopic $\Lambda'$ scenario that once the $Q\overline Q$ is
produced, they would be linked by a long string roaming inside the
detector.
The infracolor dynamics can allow the $Q \overline Q$ bound state to
survive for distances of order centimeter and preventing them from
annihilation.  In the mesoscopic $\Lambda'$ scenario, the $Q\overline
Q$ pair will appear as a single entity inside the detector.
On the other hand, if $\Lambda'$ is microscopic, the 
$Q \overline Q$ pair will lost most of their kinetic energy and angular momenta 
by emitting iglueballs and/or 
light QCD hadrons like pions before annihilation. 
In the latter case in which iquarks are QCD-colored,
it leads to a hadronic fireball along with the other SM 
decay products of the iquarkonium.
In particular, in the context of folded supersymmetry \cite{fold-susy}
it was pointed
out in Ref.\cite{Burdman:2008ek} that production of the
``squirk-antisquirk'' pair $\tilde Q {\tilde Q}^*$ at the large hadron
collider (LHC) would quickly lose their excitation energy by
bremsstrahlung and relax to the ground state of the scalar
iquarkonium. However, the energy loss due to iglueball emissions is
harder to estimate.

In this work, we consider vector-like iquarks with respect to the
electroweak gauge group but without carrying the QCD color.  However,
iquark carries a new color degree of freedom of $SU_{C^\prime}(N_{\rm
  IC})$.  Thus, iquarks do not mix with SM quarks or leptons since the
latter do not carry the new color degrees of freedom.  We
also assume MeV $\le \Lambda^\prime \ll M_Q$ so that the strings are
microscopic but yet unbreakable.  In analogous to the case of folded
supersymmetry \cite{Burdman:2008ek}, the bound states formed by the
iquark-antiiquark pairs will annihilate promptly into SM particles.
Let ${\cal Q}$ denotes a heavy iquark doublet.  The quantum number
assignment for the iquark doublet ${\cal Q}$ under
$SU_{C^\prime}(N_{\rm IC}) \times SU_C(3) \times SU_L(2) \times
U_Y(1)$ is given by
\begin{eqnarray}
{\cal Q}_{L,R} & = & \left( 
\begin{array}{c}
{\cal U}\\
{\cal D}
\end{array}
\right)_{L,R} \;\;  \sim  \;\; \left( N_{\rm IC}, 1, 2, \frac{1}{3} \right) \; .
\end{eqnarray}
These iquarks are the fractionally charged $\theta$-leptons in 
Okun's terminology \cite{thetons}.  
The gauge interactions are given by
\begin{eqnarray}
\label{lagrangian-gauge}
{\cal L}_{\rm gauge}  = 
& - & g_s^\prime G^{\prime a}_\mu \overline {\cal Q} \gamma^\mu T^a {\cal Q} 
 -   e A^\mu \left( e_{\cal U} \overline {\cal U} \gamma^\mu {\cal U} 
+ e_{\cal D} \overline {\cal D} \gamma^\mu {\cal D} \right)  \nonumber \\
& - & \frac{g}{\cos \theta_W} Z_\mu \left( v_{\cal U} \overline {\cal U} \gamma^\mu {\cal U} 
+  v_{\cal D} \overline {\cal D} \gamma^\mu {\cal D} \right)
 -  \frac{g}{\sqrt 2} \left( W^+_\mu \overline {\cal U} \gamma^\mu {\cal D} 
+ W^-_\mu \overline {\cal D} \gamma^\mu {\cal U} \right) 
\end{eqnarray}
where we have suppressed generation indices and ignored possible
mixings among iquarks.  $T^a (a = 1, \cdots, N_{\rm IC}^2 - 1)$ are
the generators of the $SU_{C^\prime}(N_{IC})$ in the defining
representation where each iquark lives and $g_s^\prime$ is its
coupling.  For vector iquark $Q = {\cal U}$ or ${\cal D}$, we have
\begin{equation}
\label{Zcouplings}
v_{Q} = \frac{1}{2} \left( T_{3}(Q_L) + T_{3}(Q_R) \right) - e_{Q} \sin^2 
\theta_W \; .
\end{equation}
Here $T_{3}(Q_{L,R})$ is the third-component of the weak isospin
for the left- (right-) handed iquark $Q$. Since we 
assume vector iquarks, $T_3(Q_L) = T_3(Q_R)$, 
they have the same value for each component of $\cal Q$.
For each vectorial iquark doublet, we can also have a Dirac bare mass term.
%
%
%
%
%
%
Due to the assumption of vectorial iquarks, precision electroweak data
from LEP do not provide any constraints even for a TeV iquark doublet.
There is no Yukawa coupling between the SM Higgs doublet and the
vector iquarks.

Iquarks in the above model can be copiously produced at the LHC, not
via normal QCD interactions, but via electroweak interactions.  The
iquarks are stable with respect to the collider time scale, and linked
by a string.  Depending on the value of $\Lambda'$ the iquarks will
lose energy by emitting iglueballs and bremsstrahlung until the iquark
pair linked by the string come together and form a iquarkonium
\cite{bj}.  The life time of the iquarkonium then depends on the
annihilation rates into SM particles.  In accord with the normal
quarkonium, the life time of iquarkonium is inversely proportional to
the square of the wave-function at the origin.  The life time and
decay patterns of the iquarkonium can therefore provide useful
information about the quantum numbers of iquarks and the dynamics of
the new strong interacting gauge group.

In this article, we set up detailed formulas for production and
decay properties of the neutral $Q\overline Q$ and charged $Q\overline Q'$
iquarkonia. 
Our result extends a previous study
on superheavy-quarkonia \cite{Barger:1987xg}
to our current interest
in iquarks and iQCD.
We will consider only one generation of iquark doublet. Extension to 
multiple generations of iquark doublets is straightforward.

The organization of the paper is as follows.  In the next section,
we present the cross sections for open production of iquarks.  
In Sec. III, we describe briefly the motion of the iquark once produced
and the radiation.  In Sec. IV, we present the results on the annihilation
of the $S$-wave iquarkonium and discuss the decay patterns. 
Conclusions are made in Sec. V.

\section{Open Uncolored Iquark Production}
Even though we assume iquarks do not carry the usual color of the strong QCD
interaction, they can still be pair produced at CERN LHC through
electroweak interactions. In the following, we list the formulas  
for the production amplitude squared of these hard partonic subprocesses.

\subsection{Open production of ${\cal U} \overline {\cal U}$ and ${\cal D}
 \overline {\cal D}$}

\subsubsection{$q (p_1) \overline q (p_2) \to {\cal U} (k_1) 
\overline {\cal U} (k_2)$ {\rm and} ${\cal D} (k_1) \overline {\cal D}(k_2)$}

There are two $s$-channel diagrams from the $\gamma$ and $Z$ exchanges 
with the following result
\begin{eqnarray}
\overline{\sum} \vert {\cal M} \vert^2 & = & 2  \, e^4  \, 
\frac{N_{\rm IC}}{3}
\left[ 
\left( \hat t - M_Q^2 \right)^2 + \left( \hat u - M_Q^2 \right)^2 + 
2 \hat s M_Q^2 
\right] \nonumber \\
& \times & 
\left[
e_q^2 e_Q^2 \frac{1}{\hat s^2} +
\frac{v_Q^2 \left( {g^q_V}^2 + {g^q_A}^2 \right) }{\sin^4\theta_W \cos^4\theta_W}
\frac{1}{\left( \hat s - M_Z^2 \right)^2 + \Gamma_Z^2 M_Z^2}  \right. \nonumber 
\\ 
& \; & \;\;\;\;\;\;\;\;\;\;\;\;\;\;  + \left.  
\frac{2e_q e_Q g^q_V v_Q}{\sin^2\theta_W \cos^2\theta_W}
\frac{1}{\hat s}
\frac{\left( \hat s - M_Z^2 \right) }{ \left( \hat s - M_Z^2 \right)^2 + 
\Gamma_Z^2 M_Z^2}
\right] 
\label{openproduction1}
\end{eqnarray}
where
$\hat s = (p_1 + p_2)^2 = (k_1 + k_2)^2$,
$\hat t = (p_1 - k_1)^2 = (p_2 - k_2)^2$ and
$\hat u = (p_1 - k_2)^2 = (p_2 - k_1)^2$;
$g_V^q = \frac{1}{2} \left( T^q_3 \right)_L - e_q \sin^2\theta_W$,
$g_A^q = \frac{1}{2} \left( T^q_3 \right)_L$ and 
$v_Q$ is given by Eq.(\ref{Zcouplings}) with $Q = {\cal U}$ or $\cal D$.
All the initial-state quark masses are set to be zero.

\subsection{Open production of ${\cal U} \overline {\cal D}$ and 
${\cal D} \overline {\cal U}$}

It turns out that the 
${\cal U}\overline{\cal D}$ and ${\cal D}\overline{\cal U}$ pairs
can be frequently produced via the virtual $W$ boson  in the
annihilation of $u {\overline d}$ and $d {\overline u}$, respectively.

\subsubsection{$u_i (p_1) \overline d_j (p_2) \to {\cal U} (k_1) 
\overline {\cal D}(k_2)$ {\rm and} $d_j \overline u_i  \to {\cal D} 
\overline {\cal U}$}

There is only one $s$-channel diagram from the $W$ exchange.
\begin{eqnarray}
\overline{\sum} \vert {\cal M} \vert^2 & = & 
\frac{ 1}{4}  \, g^4 \, \frac{N_{\rm IC}}{3}
\vert V_{ij}^{\rm CKM} \vert^2 \frac{1}{(\hat s - M_W^2)^2 + 
\Gamma_W^2 M_W^2} \nonumber\\
& \times &
\left\{
\left( \hat t - M_{\cal U}^2 \right) 
\left( \hat t - M_{\cal D}^2 \right) +
\left( \hat u - M_{\cal U}^2 \right) 
\left( \hat u - M_{\cal D}^2 \right) +
2 \hat s M_{\cal U} M_{\cal D} 
\right\}
\label{openproduction2}
\end{eqnarray}
where
$\hat s = (p_1 + p_2)^2 = (k_1 + k_2)^2$,
$\hat t = (p_1 - k_1)^2 = (p_2 - k_2)^2$ and
$\hat u = (p_1 - k_2)^2 = (p_2 - k_1)^2$.
As in the previous case, all the initial-state
quark masses are set to be zero.

\subsection{Cross sections at the LHC}

\begin{figure}[th!]
\centering
\includegraphics[width=5in]{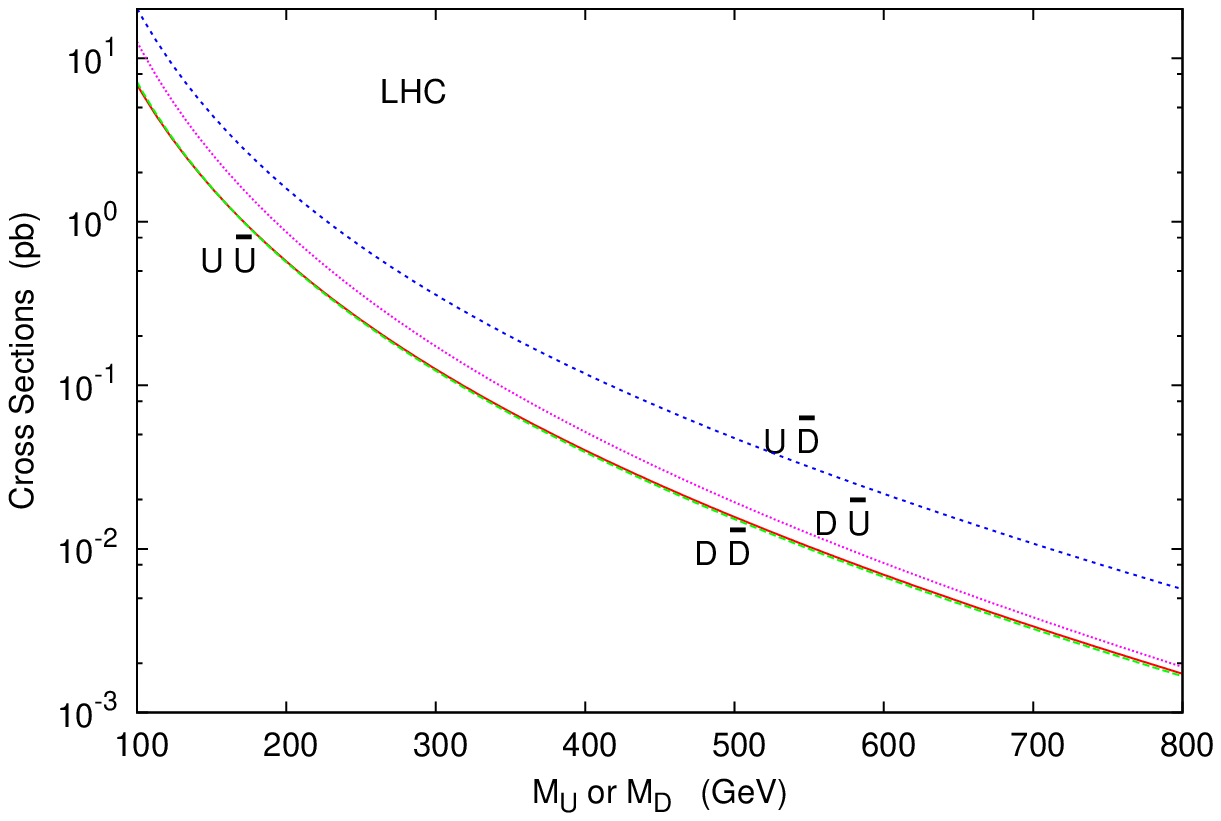}
\caption{\small \label{x-sec} Production cross sections for $pp \to
  {\cal U} \overline{\cal U},\; {\cal D} \overline{\cal D},\; {\cal U}
  \overline{\cal D}$ and ${\cal D} \overline{\cal U}$ at the LHC.  The
  label $M_{\cal U}$ on the $x$-axis is for ${\cal U} \overline{\cal
    U},\; {\cal U} \overline{\cal D}$ and ${\cal D} \overline{\cal U}$
  production while $M_{\cal D}$ is for ${\cal D} \overline{\cal D}$
  production.  We assume $M_{\cal U} - M_{\cal D} = 10$ GeV and set
  $N_{\rm IC} = 3$.  }
\end{figure}

With the above formulas
Eqs.(\ref{openproduction1})-(\ref{openproduction2}) we present in
Fig.~\ref{x-sec} the production cross sections of iquarks at the LHC.
The production rates are significant even though the iquarks are
produced via electroweak interactions rather than QCD.  For $M_{\cal
  U, D}$ around $100-200$ GeV the cross sections are of order $O(1) -
O(10)$ pb 
and are about an order of magnitude larger than the scalar iquarkonium
case in folded supersymmetry \cite{fold-susy}.
In contrast with QCD, due to the unbroken string flux
tube, the two iquarks do not hadronize individually to form isolated
jets.  Since we assume the iquark doublet is vector-like, the
$\beta$-decay between doublet members is suppressed by small mass
splittings due to radiative corrections.  Instead these initially
flying-apart iquark-antiiquark pair will come back close to each other
to form iquarkonium after losing their kinetic energies by radiating
off iglueballs and photons \cite{Kang-Luty}.
Essentially, in the case of microscopic $\Lambda'$ 
all open iquark pairs will, at the end, come together to form
a iquarkonium.  This is in sharp contrast to the normal quarkonium, in which
we have to force them to go together in the same direction (e.g. by
radiating one or more gluons) and in roughly the same velocity in order 
to form a quarkonium. Therefore, the iquarkonium production rates are 
not inferior to the quarkonium, although iquarks are only produced
via electroweak interactions.
We will discuss more about these interesting phenomena in the next section.

\section{Behavior of the open uncolored iquarks}

\subsection{Macroscopic $\Lambda'$}

The iquarks are produced via electroweak interactions.  Once they are pair
produced, they are still connected by a string or an iQCD flux tube.  Recall
in usual QCD the flux tube gets broken by creation of 
light $q\overline q$ pairs.
However, in iQCD the energy density stored in the string cannot exceed
$\Lambda'^2$, which is way smaller than the mass of iquarks.  Therefore, 
the iquark-antiiquark pair production is exponentially suppressed.  
The iglueball production from the vacuum is also expected to be suppressed
due to its finite mass gap.
Thus, the string is relatively stable.  

Nevertheless, the iquarks carry electric and weak charges so that it will
undergo bremsstrahlung and ionization with the detector materials. In 
Ref. \cite{Burdman:2008ek}, bremsstrahlung is treated semi-classically
as two massive charged particles connected a string of tension $\Lambda'^2$.
The rate energy loss is proportional to $\alpha \Lambda'^4/M^3$.  
On the other hand, the ionization energy loss when transversing the 
detector is given by the Bethe-Bloch equation \cite{pdg}. 
Essentially, the penetrating particles
lose energy by exciting the electrons of the material.
Ionization energy loss $dE/dx$
is a function of $\beta \gamma \equiv p/M$ and the charge of the
penetrating particle \cite{pdg}. 
The signature of iquarks due to the above two energy loss mechanisms is
very spectacular.  The ionization energy loss in the detector could
give rise to observable tracks in the tracking sector of the detector.
In addition, the bremsstrahlung photons may be, though soft,
identified along with the iquarks.  Since the iquarks are roaming 
around in the detector connected by a string, there may be random tracks
plus bremsstrahlung photons going along with the tracks.
At the end of kinetic-energy loss, the iquarks and antiiquarks could be still
far apart so that they cannot annihilate.  They would then stop inside the
detector.

\subsection{Microscopic $\Lambda'$}
In this case, the size of the string is so small that
when the iquarks lose their kinetic energy, they will come together to form
a iquarkonium.  The iquarkonium will then annihilate into SM particles.
We will consider the annihilation channels in the next section.  
Whether the iquarkonium can survive a period of time depends on 
(i) the annihilation rate of the iquarkonium, and
(ii) the rate of energy loss of the fast moving iquarks.
It is easy to show that the annihilation rate into SM particles is
proportional to $\alpha_W^2 |R_S(0)|^2/M^2$ 
where $\alpha_W = \alpha_{\rm em}
/\sin^2\theta_{\rm W}$, $R_S(0)$ is the radial wave-function at the origin, and 
$M$ is the mass of the iquarkonium.  For heavy iquarkonium $|R_S(0)|^2$ 
scales as $\alpha_s'^3 M^3$, so that the overall annihilation
rate scales as $\alpha_W^2 \alpha_s'^3 M$.
On the other hand, the rate of energy loss of the fast moving iquarks 
depends on the bremsstrahlung rate and the rate of radiating iglueballs.
While the rate of radiating the iglueballs is largely unknown, 
the bremsstrahlung rate is 
proportional to $\alpha_{\rm em} \Lambda'^4/M^3$.  
For $M\sim 100$ GeV the bremsstrahlung rate is faster than 
the annihilation rate.  The iquark pair loses the kinetic energy to almost
stationary and form the iquarkonium, which then survives a short period
of time before annihilates.  In this case, $P$-wave annihilation is 
negligible.  For the case of very heavy iquarks $M\sim 1$ TeV, the 
bremsstrahlung rate is comparable to the annihilation rate so that
the iquarkonium annihilates right away while the iquarks are losing the 
kinetic energy.  In this case, we expect the $P$-wave annihilation is
also relevant.  We will delay the $P$-wave annihilation
to a future publication while we focus on the $S$-wave case in this paper.
If the rate of radiating iglueball dominates over the bremsstrahlung,
the discussion here is very straight-forward.  The iquark and antiiquark
quickly loses their kinetic energy, and the size of the string is so small
that the iquark and antiiquark are readily coming together 
to form a iquarkonium, which
then annihilates promptly.   In this case, the annihilation rate is dominated by
the $S$-wave states.

\section{Iquarkonium Decay}

In this section, we present the formulas for $S$-wave iquarkonium decays. 
We have checked that some of the results are consistent with 
a previous similar calculations
for a superheavy quarkonium decays by  Barger {\it et al} \cite{Barger:1987xg} 
with appropriate modifications. 
However, parts of the results are genuinely new.

\subsection{Neutral Iquarkonium}

\subsubsection{$f \overline f$}

\begin{eqnarray}
\Gamma ( \eta_{Q \overline Q} \to f \overline f ) & = &  0 \\
\Gamma ( \psi_{Q \overline Q} \to f \overline f ) & = & \frac{4  \, 
N_{\rm IC} N_{Cf} \alpha_{\rm em}^2 \beta_f}{3} 
\left\{ \left( 1 + 2 R_f \right) \left(
e_{Q} e_f + \frac{  v_{Q} g_V^f }{ x_W \left( 1 - x_W \right) (1-R_Z)} \right)^2
\right.
\nonumber\\
&  &  \qquad + \left. 
 \beta_f^2 \frac{v_{Q}^2 {g_A^f}^2}{x_W^2 \left( 1 - x_W \right)^2}
\frac{1}{\left( 1 - R_Z \right)^2}
\right\}
\frac{\vert R_S(0) \vert^2}{M^2}
\end{eqnarray}
Here,
$N_{Cf}$ is the color factor for the fermion $f$ 
(1 for leptons and 3 for quarks),
$R_i = M_i^2 / M^2 \;\; (i = f, Z)$ with $M$ being the mass of the iquarkonium,
$x_W = \sin^2\theta_W$ and
$\beta_f = (1 - 4 R_f)^{1/2}$.

\subsubsection{$\gamma\gamma$}

\begin{eqnarray}
\Gamma ( \eta_{Q \overline Q} \to \gamma \gamma ) & = & 4 \, N_{\rm IC} 
\alpha_{\rm em}^2 e_{Q}^4
\frac{\vert R_S(0) \vert^2}{M^2} \\
\Gamma ( \psi_{Q \overline Q} \to \gamma\gamma ) & = & 0 \nonumber 
\end{eqnarray}

\subsubsection{$Z\gamma$}

\begin{eqnarray}
\Gamma ( \eta_{Q \overline Q} \to Z \gamma ) & = &8 \, N_{\rm IC} 
\alpha_{\rm em} \alpha_Z e_{Q}^2 v_{Q}^2 
\left( 1 - R_Z \right)
\frac{\vert R_S(0) \vert^2}{M^2} \\
\Gamma ( \psi_{Q \overline Q} \to Z \gamma) & = & 0 \nonumber 
\end{eqnarray}
Here, we have defined
$\alpha_Z = \alpha_{\rm em} / (\sin^2\theta_W \cos^2\theta_W)$ and
$R_Z = M_Z^2 / M^2$.

\subsubsection{$ZZ$}

\begin{eqnarray}
\Gamma ( \eta_{Q \overline Q} \to Z Z ) & = & 4 \, N_{\rm IC} 
\alpha_Z^2 v_{Q}^4 \beta_Z^3
\frac{1}{\left( 1 - 2 R_Z \right)^2}
\frac{\vert R_S(0) \vert^2}{M^2}  \\
\Gamma ( \psi_{Q \overline Q} \to Z Z) & = & 0 \nonumber  
\end{eqnarray}
where $\beta_Z = (1 - 4 R_Z)^{1/2}$.
In addition to setting the axial vector couplings of the iquarks to be zero,
we also exclude the Higgs exchange contribution in the formulas of 
Barger {\it et al} \cite{Barger:1987xg}
because a vectorial iquark does not couple to the standard model Higgs boson.

\subsubsection{$W^+W^-$}

Using the Feynman rules defined by Eq.(\ref{lagrangian-gauge}), one obtains
\begin{equation}
\Gamma (\eta_{Q \overline Q} \to W^+W^- ) =  
\frac{N_{\rm IC} \alpha_W^2 \beta_W^3}{2} 
\frac{1}{ \left( 1 - R_{Q^\prime W} \right)^2}
\frac{\vert R_S(0) \vert^2}{M^2} \; .
\end{equation}
Here $\alpha_W = \alpha_{\rm em} /\sin^2\theta_W$, 
$\beta_W = (1 - 4 R_W)^{1/2}$ and only $t$-channel exchange of the $Q'$ 
iquark contributes for the
$^1S_0$ state.
However, for the $^3S_1$ state, 
both the $t$- and $s$-channels appear with the following amplitudes
\begin{equation}
 M_t = \pm \frac{ R_S(0) \sqrt{N_{\rm IC} } } {M \sqrt{4 \pi M} } 
  \frac{2 g^2 }{1 - R_{Q'W}} \left[ (\hbox{$1\over2$} + r_{Q'} )
                                           ( k_1^\mu g^{\nu\alpha} 
         - k_2^\nu g^{\mu\alpha} )- g^{\mu\nu} k_1^\alpha \right]
 \epsilon_\alpha(P) \epsilon_\nu(k_1) \epsilon_\mu (k_2) 
 \end{equation}
where the $+$ or $-$ sign is for  the case of 
$\overline{\cal U}{\cal U}$ or $\overline{\cal D}{\cal D}$ respectively and
\begin{equation}
 M_s = -  \frac{ R_S(0) \sqrt{ N_{\rm IC}  }}  {M\sqrt{4 \pi M}} 
\frac{2 g^2 \ g_Q}{2(1-R_Z)} \left[ 
  (k_2 - k_1)^\alpha g^{\mu\nu} - 2 k_2^\nu g^{\mu\alpha} + 2 k_1^\mu g^{\nu\alpha}
 \right ]
 \epsilon_\alpha(P) \epsilon_\nu(k_1) \epsilon_\mu (k_2) \; .
\end{equation}
Here
$g_Q = (T_3 (Q_L) + T_3 (Q_R)) - 2 \, e_Q \sin^2\theta_W R_Z$, 
$r_{Q^\prime} = M_{Q^\prime} / M$ and 
$R_{Q^\prime W} = (1 - 4 R_{Q^\prime} + 4 R_W) / 2$ 
with
$(Q,Q^\prime) = ({\cal U},{\cal D})$ or $({\cal D},{\cal U})$.
Note that these two amplitudes partially cancel each other if
$r_{Q'}={1\over2} $ when the iquarks ${\cal U}$ and ${\cal D}$ are
degenerate in mass for the vectorial iquarks. The
cancellation is expected to reduce the large decay rate into
longitudinal polarizations of $W^\pm$ bosons. However, we allow the
iquark masses to be different in our formulas.
Our result of the partial width can be expressed compactly as 
\begin{equation}
 \Gamma(\psi_{Q\overline Q}\to W^+W^-)
= {N_{\rm IC} \alpha_W^2\beta^3_W \over 48 R_W^2}
  {|R_S(0)|^2 \over  M^2} 
  \left[ H^2+4 \left( H^2+3HG+G^2 \right) R_W+12G^2R_W^2 \right]  
\end{equation}
where
\begin{equation}
   G={|g_Q|\over 1-R_Z} -{1\over 1-R_{Q'W}}   \ ,\qquad
   H={|g_Q|\over 1-R_Z} -{2r_{Q'}\over 1-R_{Q'W}}   \ . 
\end{equation}

\subsubsection{$g' g'$, $g'g'g'$, $Z g' g'$, $\gamma g' g'$}

The iquarks couple to the igluon field of the $SU_{C'}(N_{IC})$ with a coupling
strength $g_s' = \sqrt{ 4 \pi \alpha_s'}$.  These igluon fields will 
escape the detection, giving rise to missing energies in the final state.
We compute the leading order of these processes. 

For the $^1S_0$ neutral quarkonium the leading decay mode involving
$g'$ is $g'g'$:
\begin{equation}
\Gamma ( \eta_{Q \overline{Q}} \to g' g' ) = \frac{ N_{IC}^2 -1}{N_{IC}}
\alpha_s'^2 \, \frac{|R_S(0)|^2}{ M^2} \;,
\end{equation}
where 
 $\alpha_s'(Q)= 12 \pi / [ (11 N_{IC} - 2 n_Q) \ln(Q^2/\Lambda'^2) ]$. 
In our numerical works presented later, 
we will choose the number of infracolor $N_{IC}=3$ and the number of iquark generation
$n_Q = 1$ at the running scale $Q=M$.

For the $^3S_1$ neutral quarkonium the leading decay modes involving $g'$ 
are $g'g'g'$, $\gamma g' g'$, and $Z g'g'$.  
The formulas for $g' g' g'$ and $\gamma g' g'$ have simple closed forms:
\begin{eqnarray}
\Gamma (\psi_{Q \overline{Q}} \to g' g'g') &=& \frac{\alpha_s'^3}{9 \pi }\, 
\frac{(N_{IC}^2-1)(N_{IC}^2-4)}{N_{IC}^2} \, \frac{ |R_S(0)|^2}{M^2}\, (\pi^2 - 9) \;, \\
\Gamma (\psi_{Q \overline{Q}} \to \gamma g'g') &=&
  \frac{4 \alpha_s'^{2} e_Q^2 \alpha}{3 \pi 
  }\, \frac{N_{IC}^2-1}{N_{IC}} \, \frac{ |R_S(0)|^2}{M^2}\, (\pi^2 - 9)\;.
\end{eqnarray}
The formula for $Z g' g'$ is shown in the appendix.

\subsection{Charged iQuarkonium}
The charged iquark-antiiquark pair once produced will settle down to
a charged iquarkonium state and finally annihilates into SM
particles. The decay formulae for each channel are listed in the
following.

\subsubsection{$f \overline f^\prime$}
\begin{equation}
\Gamma ( \psi_{{\cal U} \overline {\cal D}}  \to u_i \overline d_j )  = 
\frac{ N_{Cf} \,N_{\rm IC} \alpha_W^2   \beta_{ij} \vert 
V_{ij}^{\rm CKM} \vert^2 }{12}
\frac{1}{(1 - R_W)^2}
\left[ 2 - R_i -R_j - \left(R_i-R_j\right)^2 \right]
\frac{\vert R_S(0) \vert^2}{M^2} \;,
\end{equation}
where 
\begin{equation}
\beta_{ij} 
= \left(  \left(1-R_i-R_j \right)^2 - 4 R_i R_j \right)^{\frac{1}{2}} \;,
\end{equation}
where $R_{i,j} = r_{i,j}^2$ with $r_{i,j} = m_{i,j}/M$ and 
$M = M_{\cal U} + M_{\cal D}$.
A similar decay formula can be written down for 
$\psi_{{\cal D} \overline {\cal U}}  \to d_j \overline u_i$.

\subsubsection{$W^\pm V \; {\rm with} \; (V= \gamma , Z)$}
For the spin singlet case, we have two contributions from the $t$ and 
$u$ channels while the 
$s$ channel diagram vanishes for vector iquarks.  
\begin{eqnarray}
\label{etaplus2WplusV}
\Gamma (\eta_{{\cal U} \overline {\cal D}} \to W^+ V ) & = & 
\frac{ N_{\rm IC} \alpha_W  \alpha_V \beta_{WV}^3}{4}
\left( 
\frac{c_{\cal D}^V}{\overline r - \overline r R_W - r R_V}
+
\frac{c_{\cal U}^V}{r - r R_W - \overline r R_V}
\right)^2
\frac{\vert R_S(0) \vert^2}{M^2} \, .
\end{eqnarray}
Here we define $r = M_{\cal U}/M$, $\overline r = M_{\cal D} / M$ with 
$M = M_{\cal U} + M_{\cal D}$;
$\beta_{WV} = ((1 - R_W - R_V)^2 - 4 R_W R_V)^{1/2}$;
$c_{\cal U,D}^V = e_{\cal U,D}$ or $v_{\cal U,D}$ for $V = \gamma$ or $Z$ with 
$v_{\cal U,D}$ defined by Eq.(\ref{Zcouplings});
and
$\alpha_V = \alpha_{\rm em}$ or $\alpha_Z$ for $V = \gamma $ or $Z$ respectively.

For the spin triplet case, we have contributions from
all $s$-, $t$- and $u$-channels.
For the $W\gamma$ case, we obtain
\begin{eqnarray}
\label{psiplus2WplusGamma}
\Gamma ( \psi_{{\cal U} \overline {\cal D}} \to W^+ \gamma  ) & = & 
\frac{N_{\rm IC} \alpha_{\rm em} \,\alpha_W}{12} 
\left( \frac{e_{\cal U}}{r_{\cal U}} - \frac{e_{\cal D}}{r_{\cal D}} - 2 \right)^2
\frac{\left( 1 - R_W^2 \right)}{R_W} 
\frac{\vert R_S(0) \vert^2}{M^2}  \; .
\end{eqnarray}
It is interesting to note that all the $s$-, $t$- and $u$-channel
amplitudes of the $W^+\gamma$ mode completely cancel when $M_{\cal
  U}=M_{\cal D}$.  Such a cancellation is expected to avoid large
contributions from the longitudinal mode of the $W$ polarization.

Similarly, we work out the decay rate for the  $WZ$ case, 
\begin{eqnarray}
\label{psiplus2WplusZ}
\Gamma ( \psi_{{\cal U} \overline {\cal D}} \to W^+ Z  ) & = & 
\frac{N_{\rm IC} \alpha_W \alpha_Z \beta_{WZ}^3}{24} 
\left( X_{\cal U} - X_{\cal D} - \frac{\cos^2 \theta_W}{1 - R_W}\right)^2
\frac{1}{R_W R_Z} \frac{\vert R_S(0) \vert^2}{M^2} 
\nonumber \\
& \times & \left\{ 8 R_W + 2 R_Z \left( 1 + R_W \right) 
\left( 2 + \Delta \right)^2 
+ \left( 1 + R_W + R_Z  + R_Z \Delta  \right)^2 
\right\}
\end{eqnarray}
where we have defined
\begin{equation}
\Delta = \left( X_{\cal U} - X_{\cal D} - 
\frac{\cos^2 \theta_W}{1 - R_W}\right)^{-1}
\left(
\frac{X_{\cal U}}{r_{\cal U}} - \frac{X_{\cal D}}{r_{\cal D}} - 
2 \left( X_{\cal U} - X_{\cal D} \right)
\right)
\end{equation}
with
\begin{equation}
X_{\cal U} = \frac{v_{\cal U}}{1 - R_W - \frac{r_{\cal D}}{r_{\cal U}} R_Z} \;\;\;\;\; \mathrm{and} \;\;\;\;\;
X_{\cal D} = \frac{v_{\cal D}}{1 - R_W - \frac{r_{\cal U}}{r_{\cal D}} R_Z}   \; .
\end{equation}
%
%
%
We have used a relation:
\begin{equation}
e_{\cal U} \left( \frac{r_{\cal D}}{r_{\cal U}} - 1 \right) - e_{\cal D} 
\left( \frac{r_{\cal U}}{r_{\cal D}} - 1 \right) =
\frac{e_{\cal U}}{r_{\cal U}} - \frac{e_{\cal D}}{r_{\cal D}} - 2 
\end{equation}
due to the identity $e_{\cal U} - e_{\cal D} - 1 = 0$ as a consequence of 
charge conservation. 
We take $e_{\cal U} = 2/3$ and $e_{\cal D}=-1/3$ as
implied by our hypercharge assignment.

\noindent
The CP conjugate processes give the same widths by symmetry, viz.
$$ \Gamma(\eta_{{\cal D} \overline {\cal U}} \to W^- V)
=  \Gamma(\eta_{{\cal U} \overline {\cal D}} \to W^+ V) \ ,\quad
   \Gamma(\psi_{{\cal D} \overline {\cal U}} \to W^- V)
=  \Gamma(\psi_{{\cal U} \overline {\cal D}} \to W^+ V)  \ .
$$

\subsubsection{$W g' g'$}
For charged $^1S_0$ iquarkonium the decay into $W g' g'$ is zero 
on the amplitude level in the
limit of degenerate $M_{\cal U} = M_{\cal D}$.  So we ignore this mode
in the decay branching ratio of the charged $^1S_0$ iquarkonium.
On the other hand, the leading mode for the charged $^3S_1$ iquarkonium is 
nonzero, and we list the formulas in the appendix.  We include
the $W g'g'$ in the decay branching ratio.

\subsection{Decay patterns}

\begin{figure}[bh!]
\centering
\includegraphics[width=3.2in]{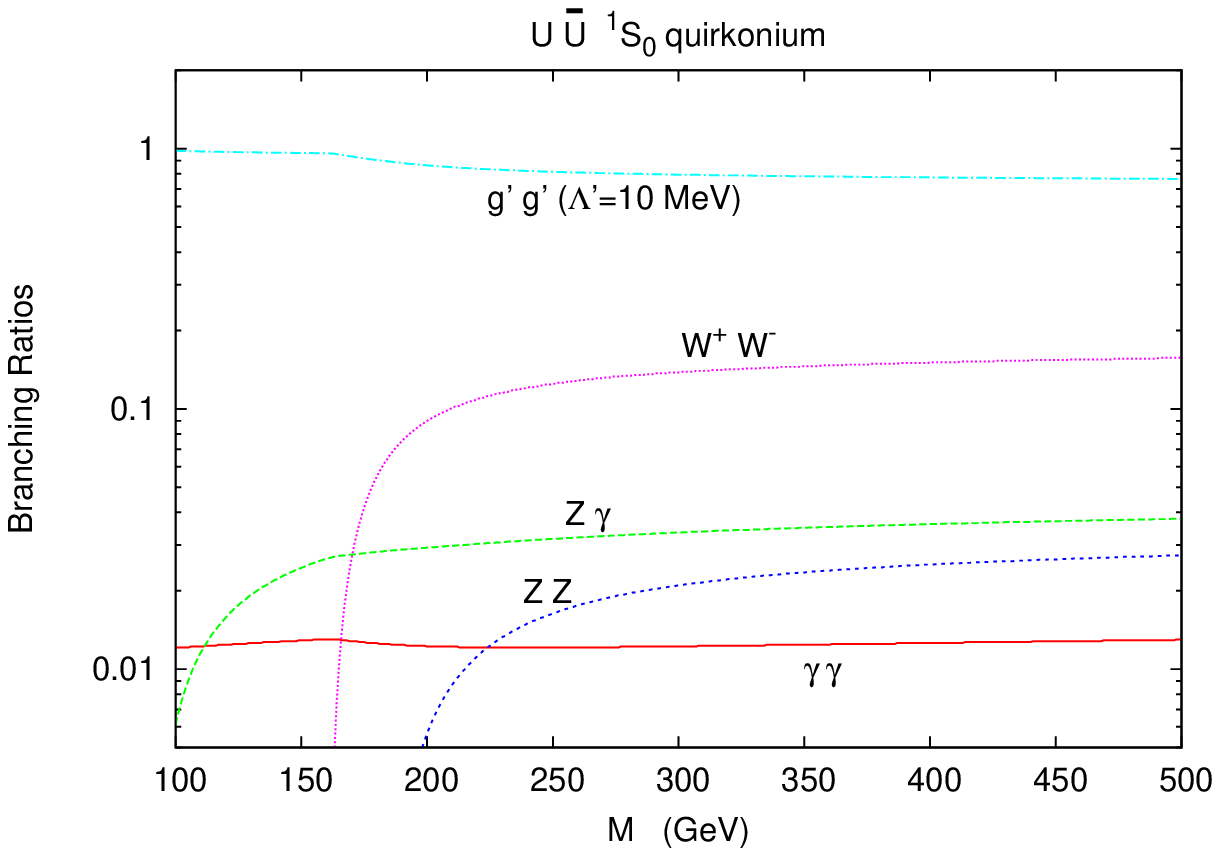}
\includegraphics[width=3.2in]{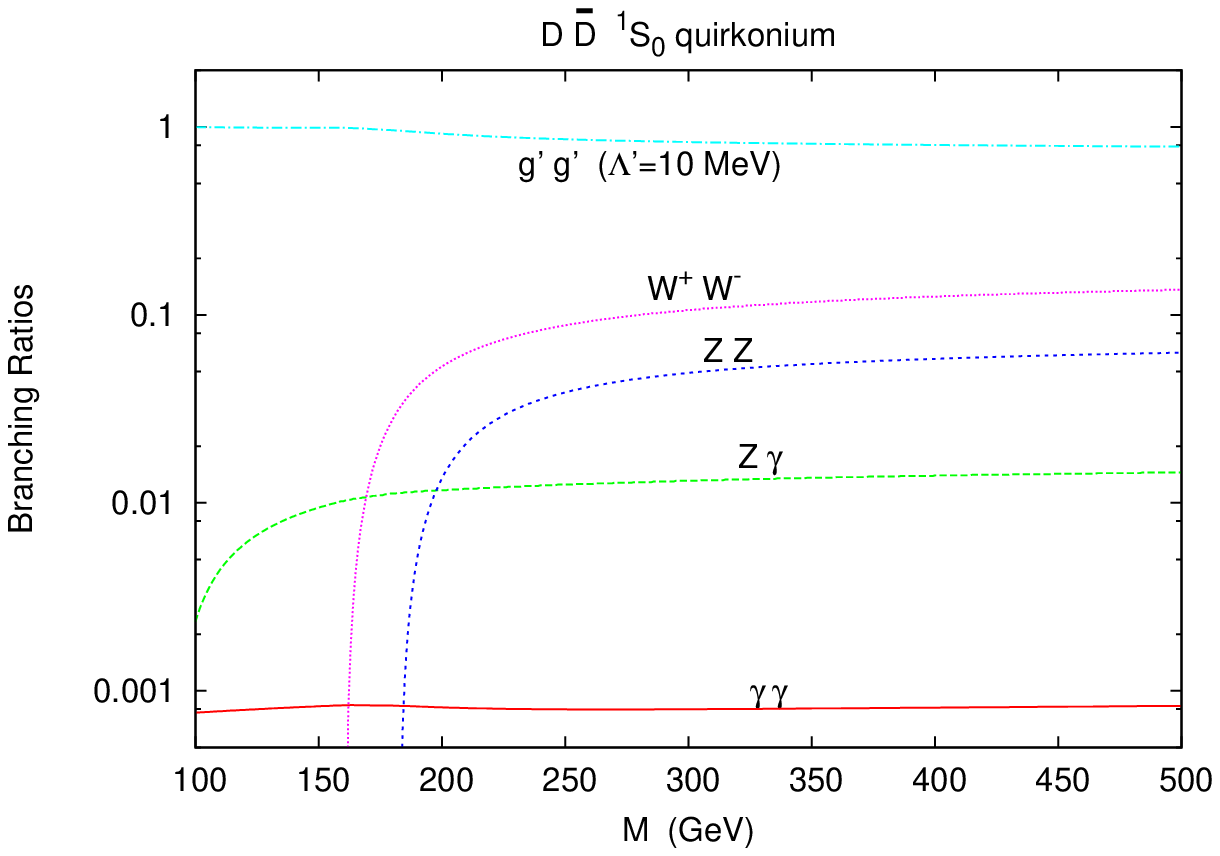}
\caption{\small \label{fey1}
Branching fractions of the iquarkonium of
(a) $^1 S_0 ({\cal U}\overline{\cal U})$  and 
(b) $^1S_0({\cal D}\overline{\cal D})$ 
versus the iquarkonium mass $M$.
We have chosen $\Lambda'=10$ MeV and $n_Q=1$ in the running of $\alpha_s'$.}
\end{figure}

\begin{figure}[th!]
\centering
\includegraphics[width=3.2in]{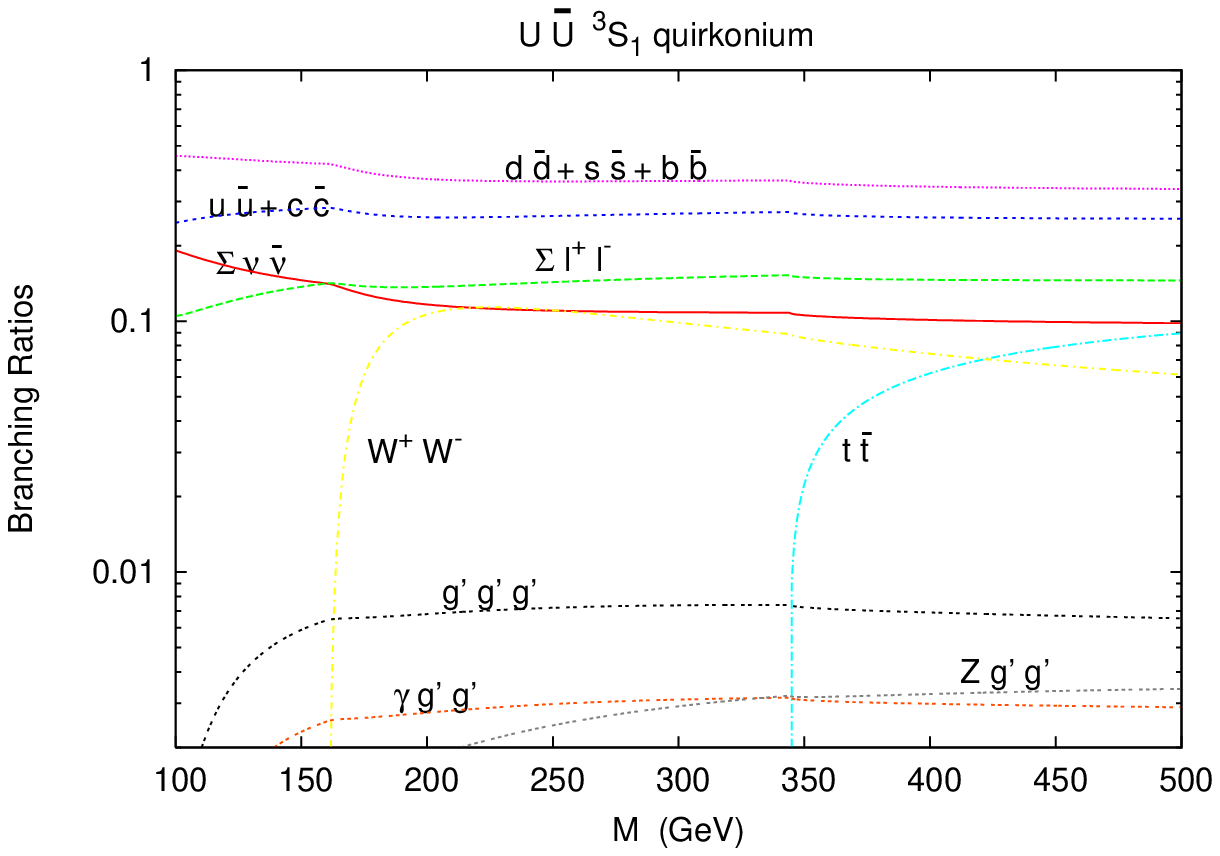}
\includegraphics[width=3.2in]{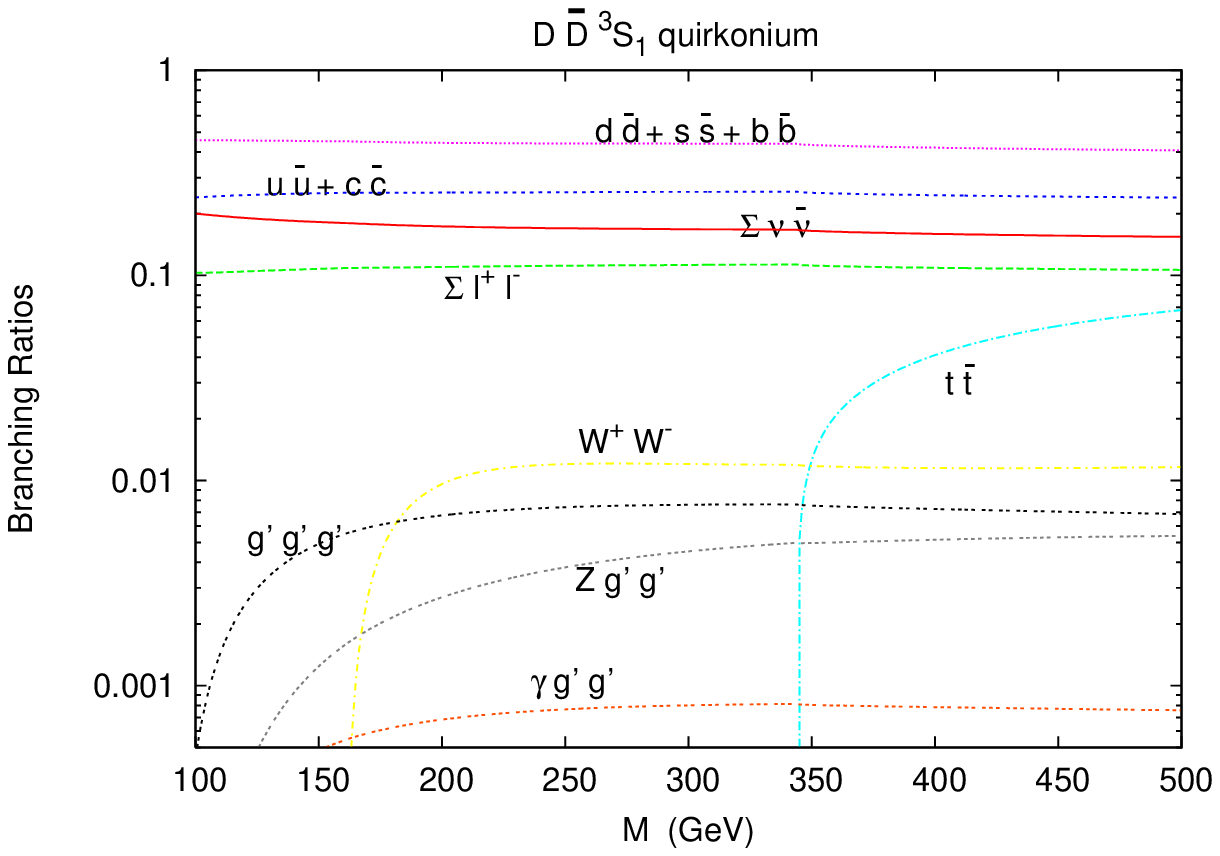}
\caption{\small \label{fey2}
Branching fractions of the iquarkonium of
(a) $^3S_1({\cal U}\overline{\cal U})$ and 
(b) $^3S_1({\cal D}\overline{\cal D})$ 
 versus the iquarkonium mass $M$.
We have chosen $\Lambda'=10$ MeV and $n_Q=1$ in the running of $\alpha_s'$.}
\end{figure}

\begin{figure}[t!]
\centering
\includegraphics[width=3.2in]{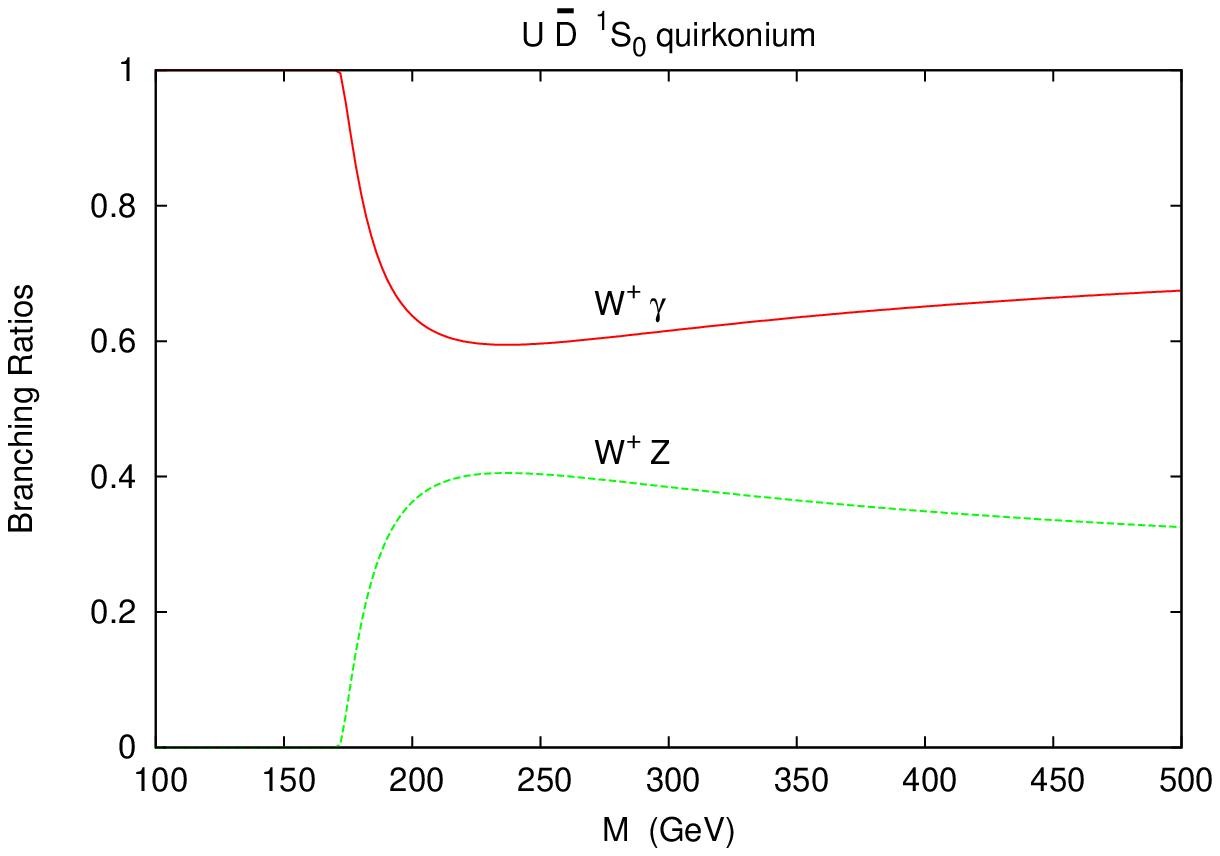}
\includegraphics[width=3.2in]{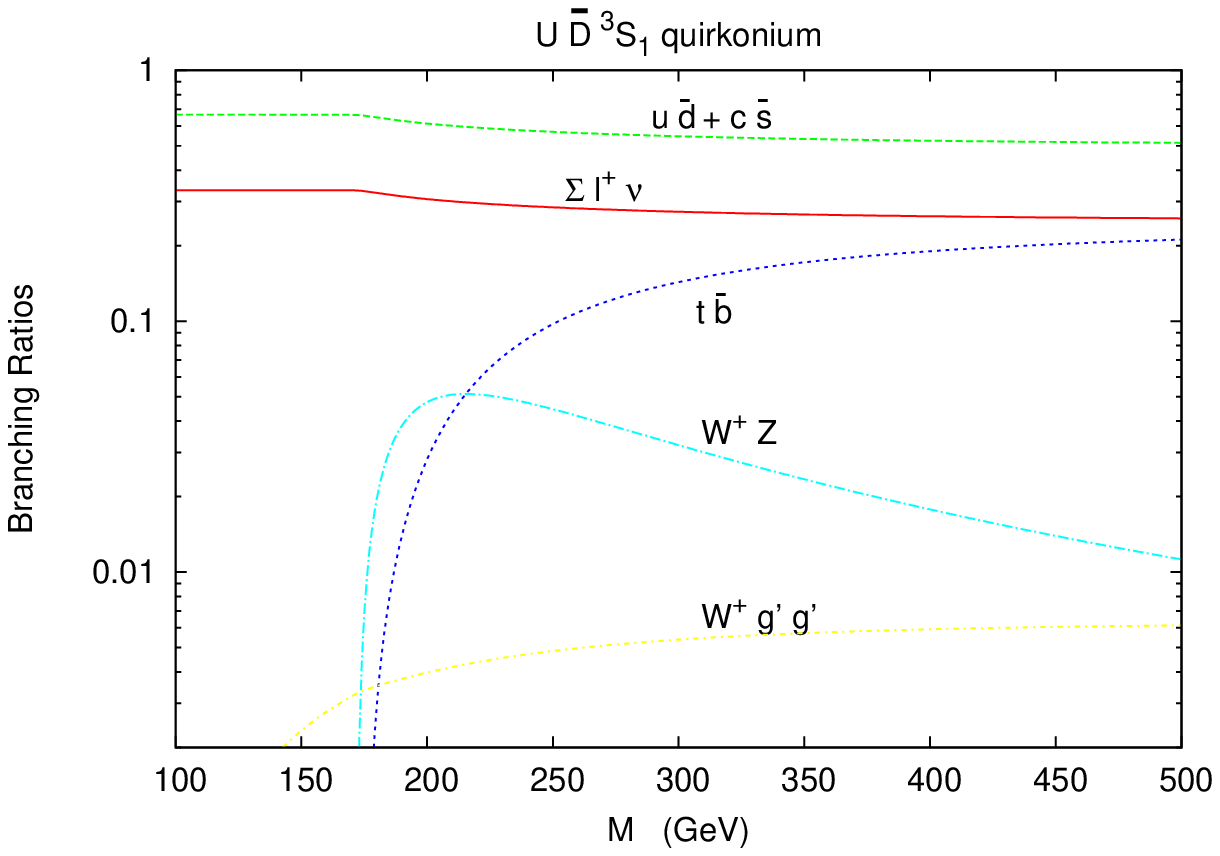}
\caption{\small \label{fey3}
Branching fractions of the charged iquarkonium of
(a) $^1S_0({\cal U}\overline{\cal D})$ and 
(b) $^3S_1({\cal U}\overline{\cal D})$ 
 versus the iquarkonium mass $M$.
We have chosen $\Lambda'=10$ MeV and $n_Q=1$ in the running of $\alpha_s'$.}
\end{figure}

We present the decay branching ratios of the $S$-wave $^1S_0$ and 
$^3S_1$ iquarkonium of 
${\cal U}\overline {\cal U}$, ${\cal D}\overline {\cal D}$, 
and the charged ${\cal U}\overline {\cal D}$ in Figs.~\ref{fey1} -- \ref{fey3}. 
In these plots, we have set $N_{\rm IC} = 3$ and a small mass difference of
$M_{\cal U} - M_{\cal D} = 10$ GeV for
neutral and charged iquarkonium.

The pseudoscalar state, $\eta_{Q\overline{Q}}$, 
only decays into a pair of gauge bosons
for both ${\cal U}\overline {\cal U}$ and ${\cal D}\overline {\cal D}$,
as shown in Fig. \ref{fey1}.
The dominant decay mode for $\eta_{Q\overline{Q}}$ is $g' g'$, which is 
still valid for $\Lambda'$ down to 1 MeV. It gives rise to an invisible decay.
The second largest decay mode is  $W^+ W^-$ 
when the mass $M$ of the iquarkonium is above $2 m_W$ threshold;
otherwise $\gamma\gamma$ and $Z\gamma$ are large when 
$M$ of the iquarkonium is below $2 m_W$ threshold.
On the other hand, the fermion-antifermion modes are dominant in the decay of
$\psi_{Q\overline{Q}}$ states; especially the down-type quarks, 
followed by the up-type quarks and then lepton modes: see  Fig.~\ref{fey2}. 
The branching ratio into $W^+ W^-$ is small and so are the 
$Z g' g'$, $\gamma g' g'$, and $g'g'g'$ modes.
We show the decay branching ratios for the charged 
$\eta_{ {\cal U}\overline{\cal D}}$ and $\psi_{ {\cal U}\overline{\cal D}}$
in Fig.~\ref{fey3}.
There are only two modes for 
$\eta_{ {\cal U}\overline{\cal D}}$ state, namely $W^+\gamma$
and $W^+Z$. 
Whereas for the  $\psi_{ {\cal U}\overline{\cal D}}$
state the fermion pair modes dominate.

\subsection{Comparison with 4-th generation quarkonium}

A superheavy quarkonium made up of a pair of 
sequential 4-th generation quark and
antiquark has QCD as well as Higgs interactions.  
We have to assume that each of the 4-th
generation quark is stable against weak decay. In general it is not true, 
but just for comparison with iquarkonium we temporarily assume this is actually the case.
The major decay mode for the $^1S_0$ state
is $gg$ while that of $^3S_1$ state is $ggg$.  These QCD decays are far 
more efficient than those of electroweak decays.  Therefore, we can make the
following observations

\begin{enumerate}
\item 

The major decay mode of a $\eta_{Q\overline{Q}}$ iquarkonium 
is $g'g'$, followed by $\gamma\gamma, Z\gamma,
WW$, and $ZZ$ so that it decays mainly into an invisible mode, which is 
very different from the 2-jet mode of a superheavy quarkonium.

 \item The majority of the decay modes of a $\psi_{Q\overline{Q}}$ 
iquarkonium for all its mass range is $d\bar d + s\bar s + b\bar b$, 
which give rises to two jets in the final state. 
On the other hand, once the $WW$ channel is open, it will dominate over all other modes
in the superheavy quarkonium decay while this mode occupies only about 
1\% for iquarkonium decay.
%
Presumably the superheavy quarkonium decays dominantly into the longitudinal
components of the $W$ bosons, while
for the vectorial iquarks the longitudinal piece is largely cancelled 
in the limit of degenerate $M_{\cal U}= M_{\cal D}$ and $M 
\gg m_W$.
Furthermore, the $3g$ mode of the superheavy quarkonium will give rise to a 3-jet
final state while the $3g'$ mode of the iquarkonium has small branching ratio 
and is invisible.

 \item For the $^1S_0$ charged quarkonium it decays into 
$W\gamma$ or $WZ$ in the leading order.
This is similar to $\eta_{ {\cal U} \overline{\cal D}}$, so it is hard to
distinguish the charged $^1S_0$ state.

 \item The $^3S_1$ charged quarkonium will decay into fermion pairs via 
a virtual $W$ boson, which is similar to the charged iquarkonium.  However,
the $^3S_1$ charged quarkonium also has the $Wgg$ mode, which is
comparable to its $f\bar f'$ mode, while the $W g'g'$ mode of
the iquarkonium is very small.  So, there is a chance to distinguish
between the $^3S_1 $ charged iquarkonium from the quarkonium.
\end{enumerate}

\section{Conclusions}

Production of iquark-antiiquark pair connected by unbreakable long and
stable string associated with a new confining non-abelian gauge group
may lead to spectacular events at the LHC.  We presented in some
details the phenomenology of uncolored iquarks in this work.  Since
the momentum transfer for the electroweak processes producing the
iquark-antiiquark pairs is typical of order $M_Q$ and one has to bring
the pairs at a distance of order $1/M_Q$ in order for them to
annihilate, the bound states are thus formed with high excitation
energies and large orbital momenta.  These energies are quickly
dissipated by emitting iglueballs and photons before they settle down
in the ground state and annihilate into standard model particles.

We have studied open production of the uncolored iquarks at the LHC
and the leading-order 2-body and 3-body decays of
$S$-wave iquarkonium.  The decay patterns of $S$-wave iquarkonium formed
by a vector-like iquark doublet are found to be distinguishable from a
superheavy quarkonium composed of a sequential 4-th generation quark
doublet.
With our choice of $\Lambda' \ \sim \ 10$ MeV, 
the emitted iglueballs due to energy loss as the iquark-antiiquark crossing each other
before annihilation 
may decay outside the detectors and become invisible
\cite{Kang-Luty}.
Developing new search strategies at the LHC for
detecting the decay products of iquarkonium together with many soft
photons emitted due to the energy loss are very important to unravel
this kind of new physics beyond the standard model.
Some of these issues have been addressed in a recent article of Ref.\cite{Harnik-Wizansky}.
For cosmological implications of iQCD, we refer the readers to the literature \cite{quirky-cosmology}.

\vspace{.5cm} \acknowledgments This work was supported in part by the
NSC under grant No. NSC 96-2628-M-007-002-MY3, the NCTS of Taiwan, the
Boost Program of NTHU, as well as U.S. DOE under grant
No. DE-FG02-84ER40173.  WYK would like to thank the support from the
visitor program of NCTS where this work was initiated.  TCY would like
to thank the hospitality from the Kavli Institute for Theoretical
Physics China at the Chinese Academy of Sciences and the Physics
Department of Peking University where this work was partly performed.
KC would like to thank the hospitality of the Institute of Theoretical
Physics at the Chinese University of Hong Kong.

\appendix
\section{Three-body decay formulas}

For the process $\psi_{{\cal U} \overline {\cal D}} \to W^+ g' g'$
we define the rescaled energy variables in the rest frame of
the iquarkonium:
\[
  x_1 = \frac{2 E_1}{M}\,, \qquad x_2 = \frac{2 E_2}{M} \,, \qquad 
 x_v  = \frac{2 E_W}{M}
\]
so that $x_1 + x_2 + x_v=2$. We also define 
\[
  \xi = \frac{m_W^2}{M^2} \; .
\]
The differential width, in the limit of degenerate $M_{\cal U}=M_{\cal D}$,
is given by
\begin{eqnarray}
\frac{d \Gamma}{d x_v d x_1} (\psi_{ {\cal U} \overline{\cal D}} \to W^+ g' g')
&=& \frac{2 \alpha_s'^{2} \alpha}{3 \pi 
  \sin^2\theta_{\rm w} }\, \frac{N_{IC}^2-1}{N_{IC}} \, \frac{ |R_S(0)|^2}{M^2}\,
 \frac{1}{x_1^2 x_2^2 (x_v - 2 \xi)^2 }\, \nonumber \\
 && \hspace{-1in} \times 
\biggr [ 
2\xi^4 + 2\xi^3 (6 -4 x_v +2 x_1 -x_v x_1 -x_1^2 ) \nonumber \\
&& \hspace{-0.9in}
+  2\xi^2 \left( 11-16x_v +6x_v^2 - (8 -2x_v  -x_v^2) x_1 + (4 +x_v) x_1^2
  \right )  \nonumber \\
&& \hspace{-0.9in}
+  \xi \biggr( 4(1-x_v)(4-5x_v+2x_v^2)  - (32 -44 x_v 
   + 14 x_v^2) x_1 + (20  -18 x_v +x_v^2 ) x_1^2 \nonumber \\
&& \qquad - 2(2 - x_v) x_1^3 +x_1^4       \biggr) \nonumber \\
&& \hspace{-0.9in}
+ 2 \biggr( 2 - 6x_v +7x_v^2 -4x_v^3 +x_v^4 - (6 -13 x_v +9 x_v^2  -
 2 x_v^3 )x_1  \nonumber \\
&& \qquad + (7 -9x_v  +3 x_v^2 )x_1^2 - 2(2 - x_v) x_1^3 +x_1^4 \biggr) \biggr ] \; .
\end{eqnarray}
The ranges of integration for $x_v$ and $x_1$ are  
\begin{eqnarray}
2\sqrt{\xi} \le & x_v & \le 1+\xi   \; , \\
\frac{1}{2}\left(2-x_v -\sqrt{x_v^2 -4\xi} \right ) \le & x_1 &  \le
\frac{1}{2}\left(2-x_v +\sqrt{x_v^2 -4\xi} \right ) \; .
\end{eqnarray}
Note that $\eta_{{\cal U} \overline {\cal D}} \to W^+ g' g'$ is
zero on the amplitude level 
in the limit of degenerate $M_{\cal U}=M_{\cal D}$.  

The substitutions needed to obtain 
$\psi_{{\cal U} \overline {\cal U}} \to Z g' g'$ or 
$\psi_{{\cal D} \overline {\cal D}} \to Z g' g'$ from the above
formula are
\[
 \frac{\alpha}{2\sin^2\theta_{\rm w}} 
\longrightarrow \frac{\alpha v^2_Q}{\cos^2\theta_{\rm w} \sin^2\theta_{\rm w}}
\;, \qquad
  m_W \longrightarrow m_Z \;.
\]
The decay of $\psi_{{\cal U} \overline{\cal U}} \to \gamma g' g'$ or
$\psi_{{\cal D} \overline{\cal D}} \to \gamma g' g'$ can be easily obtained from
the above formula with the replacement
\[
   \frac{\alpha}{2 \sin^2\theta_{\rm w}} 
  \longrightarrow e_Q^2 \alpha \,, \qquad \xi \longrightarrow 0 \; .
\]
So the differential partial width is given by
\begin{eqnarray}
\frac{d \Gamma}{d x_v d x_1} (\psi_{Q\overline{Q}} \to \gamma g'g')
   &=& \frac{4 \alpha_s'^{2} \alpha e_Q^2}{3 \pi 
  }\, \frac{N_{IC}^2-1}{N_{IC}} \, \frac{ |R_S(0)|^2}{M^2}\,
 \frac{1}{x_1^2 x_2^2 (x_v - 2 \xi)^2 }\, \nonumber \\
& \times & \biggr [ 
2 \biggr( 2 - 6x_v +7x_v^2 -4x_v^3 +x_v^4 
 - (6 -13 x_v +9 x_v^2  -
 2 x_v^3 )x_1 
 \nonumber \\
&& + (7 -9x_v  +3 x_v^2 )x_1^2 
 - 2(2 - x_v) x_1^3 +x_1^4 \biggr) \biggr ] \;, \nonumber \\
\end{eqnarray}
with the integration range
\[
0 \le  x_v  \le 1,\;\;\;\;\;    1-x_v \le  x_1  \le 1 \;.
\]
After integrating over $x_1$ and $x_v$ we obtain
\begin{equation}
\Gamma (\psi_{Q\overline{Q}} \to \gamma g'g') 
= \frac{4 \alpha_s'^{2} e_Q^2 \alpha}{3 \pi 
  }\, \frac{N_{IC}^2-1}{N_{IC}} \, \frac{ |R_S(0)|^2}{M^2}\, (\pi^2 - 9) \;.
\end{equation}
Finally, the decay width of $\psi_{Q\overline{Q}} \to g' g' g' $ is 
\begin{equation}
\Gamma (\psi_{Q\overline{Q}} \to g' g'g') = \frac{\alpha_s'^3}{9 \pi }\, 
\frac{(N_{IC}^2-1)(N_{IC}^2-4)}{N_{IC}^2} \, \frac{ |R_S(0)|^2}{M^2}\, (\pi^2 - 9) \;.
\end{equation}



\begin{thebibliography}{}
\bibitem{thetons}
L.~B.~Okun,
  JETP Lett.\  {\bf 31}, 144 (1980)
  [Pisma Zh.\ Eksp.\ Teor.\ Fiz.\  {\bf 31}, 156 (1979)];
Nucl.\ Phys.\  B {\bf 173}, 1 (1980).

\bibitem{bj}
J. D. Bjorken, SLAC-PUB-2372 (1979), 
in {\it Quantum Chromodynamics}, proceedings of the SLAC Summer Institute on Particle Physics,
Stanford, California, 1979, edited by Anne Mosher (SLAC, Stanford, 1980).

\bibitem{Gupta-Quinn}
S. Gupta and H. R. Quinn, 
Phys. Rev. D {\bf 25}, 838 (1982).


\bibitem{Kang-Luty}
  J.~Kang and M.~A.~Luty,
  arXiv:0805.4642 [hep-ph].
\bibitem{Strassler:2006im}
M.~J.~Strassler and K.~M.~Zurek,
Phys.\ Lett.\ B {\bf 651}, 374 (2007)
[arXiv:hep-ph/0604261].


\bibitem{fold-susy}
G.~Burdman, Z.~Chacko, H.~S.~Goh and R.~Harnik,
JHEP {\bf 0702}, 009 (2007) [arXiv:hep-ph/0609152].


\bibitem{Burdman:2008ek}
  G.~Burdman, Z.~Chacko, H.~S.~Goh, R.~Harnik and C.~A.~Krenke,
  Phys. Rev. D {\bf 78}:075028 (2008),
  arXiv:0805.4667 [hep-ph].

\bibitem{Barger:1987xg}
  V.~Barger, E.~W.~N.~Glover, K.~Hikasa, W.-Y.~Keung, M.~G.~Olsson,
  C.~J.~Suchyta III and X.~R.~Tata,
  Phys.\ Rev.\  D {\bf 35}, 3366 (1987)
  [Erratum-ibid.\  D {\bf 38}, 1632 (1988)].

\bibitem{pdg}
Particle Data Group, J. Phys. G {\bf 33}, 1 (2006).

\bibitem{Harnik-Wizansky}
R. Harnik and T. Wizansky,
arXiv:0810.3948 [hep-ph].

\bibitem{quirky-cosmology}
J. Kang, M. A. Luty and S. Nasri,
JHEP {\bf 0809}:086 (2008) [arXiv:hep-ph/0611322];
C. Jacoby and S. Nussinov, arXiv:0712.2681 [hep-ph].


\end{thebibliography}
\end{document}